\documentclass[preprint,12pt]{elsarticle}

\usepackage{amssymb}

\usepackage{array}
\usepackage{comment}

\journal{Computer Physics Communications}

\begin{document}

\begin{frontmatter}



\title{Numerical simulation code for self-gravitating Bose-Einstein condensates}


\author[1]{Enik\H{o} J. M. Madarassy}
\author[2]{Viktor T. Toth}

\address[1]{Division of Astronomy and Space Physics, Uppsala University, 751 20 Uppsala, Sweden}
\address[2]{Ottawa, Ontario, K1N 9H5 CANADA}


\begin{abstract}
We completed the development of simulation code that is designed to study the behavior of a conjectured dark matter galactic halo that is in the form of a Bose-Einstein Condensate (BEC). The BEC is described by the Gross-Pitaevskii equation, which can be solved numerically using the Crank-Nicholson method. The gravitational potential, in turn, is described by Poisson's equation, that can be solved using the relaxation method. Our code combines these two methods to study the time evolution of a self-gravitating BEC. The inefficiency of the relaxation method is balanced by the fact that in subsequent time iterations, previously computed values of the gravitational field serve as very good initial estimates. The code is robust (as evidenced by its stability on coarse grids) and efficient enough to simulate the evolution of a system over the course of $10^{9}$ years using a finer ($100\times 100\times 100$) spatial grid, in less than a day of processor time on a contemporary desktop computer.

\end{abstract}


\end{frontmatter}




{\bf PROGRAM SUMMARY
}

\begin{small}
\noindent
{\em Manuscript Title:} Numerical simulation code for self-gravitating Bose-Einstein condensates\\
{\em Authors:} {Enik\H{o} J. M. Madarassy} and {Viktor T. Toth}\\
{\em Program Title:} bec3p\\
{\em Journal Reference:}                                      \\
{\em Catalogue identifier:}                                   \\
{\em Licensing provisions:} none                              \\
{\em Programming language:} C++ or FORTRAN\\
{\em Operating system:} Linux or Windows                                      \\
{\em Number of processors used:} 1                             
{\em Classification:} 1.5 Relativity and Gravitation                                        \\
{\em Nature of problem:}\\
  Simulation of a self-gravitating Bose-Einstein condensate by simultaneous solution of the Gross-Pitaevskii and Poisson equations in three dimensions.
   \\
{\em Solution method:}\\
The Gross-Pitaevskii equation is solved numerically using the Crank-Nicholson method; Poisson's equation is solved using the relaxation method. The time evolution of the system is governed by the Gross-Pitaevskii equation; the solution of Poisson's equation at each time step is used as an initial estimate for the next time step, which dramatically increases the efficiency of the relaxation method.
   \\
{\em Running time:}\\
  Depends on the chosen size of the problem. On a typical personal computer, a $100\times 100\times 100$ grid can be solved with a time span of 10~Gyr in approx. a day of running time.
   \\

\end{small}

\section{Introduction}
\label{}

The rotation of spiral galaxies does not follow simple predictions based on Newton's laws. Instead, the rotational velocity curve of most spiral galaxies, plotted as a function of radial distance from the galaxy center, remains ``flat'' for a broad range of radii. The standard proposal to resolve this problem is to presume the existence of a ``dark matter halo'', which contains most of the mass of a spiral galaxy. To maintain consistency with the predictions of the most broadly accepted cosmological models, this halo must necessarily consist of ``exotic'' matter, i.e., matter predominantly composed of something other than baryons. The halo must also be collisionless and not interacting with baryonic matter \cite{Weinberg2008}.

The existence of such a halo with a suitable geometry can account for the observed rotation curves of visible matter. However, a difficult problem is to construct a dark matter halo that is gravitationally stable and does not predict excessive dark matter densities in the inner parts of the galaxy where most visible matter resides. This issue is known as the ``cuspy halo problem'' in the relevant literature \cite{2010AdAst2010E...5D}.

A recent proposal \cite{1994PhRvD..50.3650S,1994PhRvD..50.3655J,2000PhRvL..85.1158H,Sahni2000,2007JCAP...06..025B,2011PhRvD..84d3532C} addresses the cusp problem by a dark matter halo that forms a Bose-Einstein condensate (BEC) \cite{1924ZPhy...26..178B,Einstein1925}.
A particularly intriguing argument is that the condensate dark matter is, in fact, axions \cite{2012arXiv1210.0040S}.
The dynamics of a BEC halo may be determined by the balance of the attractive force of gravity and a repulsive effective long-range interaction \cite{Khlopov1985,Khlopov2000,Khlopov2002,Khlopov2005} (see also \cite{Khlopov2004}).
In particular, as the dark matter halo dominates the gravitational field of a spiral galaxy in its outer regions, a simulation that is restricted to just the halo should be sufficient to determine if a field can be obtained that yields the desired circular orbital velocities.

In the present paper, we discuss a simulation tool that we constructed to explore the dynamics of a galactic BEC halo.
The tool is not intended in its present form to study the core-cusp problem; however, we anticipate that it will be useful for investigating the rotational velocities of a galaxy surrounded by a BEC halo.
Our work is based primarily on our previous simulation of BEC in laboratory conditions, described by the non-linear Schr\"odinger equation, also known in the literature as the Gross-Pitaevskii equation. Whereas in the laboratory, a BEC characterized by a repulsive interaction is held together by an artificially introduced trapping potential, in the case of a galaxy floating in empty space, the trapping potential must be replaced by self-gravity. A numerical solution must, therefore, simultaneously address the initial value problem of the Gross-Pitaevskii equation and the boundary condition problem of Poisson's equation \cite{2012JCAP...02..011L}.

In Sec.~\ref{sec:GPE}, we introduce the dimensionless form of the Gross-Pitaevskii equation used in our computations, and the method used to solve this equation efficiently. In Sec.~\ref{sec:PE} we discuss Poisson's equation for gravity and the relaxation method. In Sec.~\ref{sec:UNITS} we elaborate on the use of physical units that are suitable for such a simulation in an astrophysical context. The problem of using suitable initial conditions to form a stable halo is briefly discussed in Sec. \ref{sec:INIT}. In Sec.~\ref{sec:CODE} we discuss the implementation of our method in FORTRAN and C++, and also comment on the possible use of GPUs for accelerated computation. Finally, our conclusions and outlook are presented in Sec.~\ref{sec:END}.

\section{Solving the Gross-Pitaevskii equation}
\label{sec:GPE}

A self-interacting, optionally rotating Bose-Einstein condensate is described accurately by a form of the time-dependent nonlinear Schr\"odinger equation known as the Gross-Pitaevskii equation \cite{springerlink:10.1007/BF02731494,Pitaevskii1961}. For computational purposes, it is advantageous to use a dimensionless form of this equation, which takes the form \cite{2006PhRvL..97x0404K}:
\begin{equation}
(i-\gamma)\frac{\partial\psi}{\partial t}=\hat{H}\psi,\label{eq:GPE}
\end{equation}
where $\gamma$ is a softening parameter
that may also be viewed as a phenomenological parameter characterizing dissipation
($\gamma=0$ is a valid choice), $\psi$ is the wave function, $t$ is time, and $\hat{H}$ is the Hamilton-operator, which in turn is given by
\begin{equation}
\hat{H}=-\frac{1}{2}\nabla^2+V.
\end{equation}
The potential $V$ is the sum of classical potentials (e.g., gravitational potential, trapping potential), the chemical potential, the non-linear term, and a rotational term:
\begin{equation}
V=\phi+\kappa|\psi|^2-\mu-\Omega L_z,\label{eq:V}
\end{equation}
where $\kappa$ represents the interaction strength, $\Omega$ is the angular velocity, and $L_z=i(x\partial_y-y\partial_x)$. We assume that the condensate's net rotation is in the $x-y$ plane.

In earlier work \cite{2008JLTP..152..122M,2009GApFD.103..269M,2009CaJPh..87.1013M,Madarassy2010}, we solved the Gross-Pitaevskii equation
numerically using the Crank-Nicholson method in combination with Cayley's formula \cite{NUMRC}, in the presence of an isotropic trapping potential (for a numerical solution in the presence of an anisotropic trap, see \cite{Muru2009,Muru2012}). In particular, the use of Cayley's formula ensures that the numerical solution remains stable and the unitarity of the wavefunction is maintained.

The value $\psi_{n+1}$ of the wavefunction at the $(n+1)$-th time step is obtained from the known values $\psi_n$ at the $n$-th time step by solving the following equation:
\begin{equation}
\left(1+\frac{1}{2}i\Delta t\hat{H}\right)\psi^{n+1}=\left(1-\frac{1}{2}i\Delta t\hat{H}\right)\psi^n.\label{eq:CrankNich}
\end{equation}
After evaluating the right-hand side given $\psi_n$, the left-hand side can be solved for. If $\hat H$ is a linear operator, this is a linear system of equations for the unknown values $\psi_{n+1}$.

In the one-dimensional case, the Hamilton operator reads
\begin{equation}
\hat{H}=-\frac{1}{2}\frac{\partial^2}{\partial x^2}+V.
\end{equation}
The second derivative can be approximated as a finite difference:
\begin{equation}
\frac{\partial^2\psi}{\partial x^2}=\frac{\psi_{k-1}-2\psi_k+\psi_{k+1}}{(\Delta x)^2},
\end{equation}
Substituting this into Eq.~(\ref{eq:CrankNich}), we obtain
\begin{eqnarray}
&&\left[1-\frac{i\Delta t}{2}\left(V+\frac{1}{(\Delta x)^2}\right)\right]\psi^{n+1}_k
-\frac{i\Delta t}{4(\Delta x)^2}\left(\psi^{n+1}_{k-1}+\psi^{n+1}_{k+1}\right)\qquad\qquad\qquad\nonumber\\
&&\qquad\qquad{}=\left[1-\frac{i\Delta t}{2}\left(V+\frac{1}{(\Delta x)^2}\right)\right]\psi^n_k
+\frac{i\Delta t}{4(\Delta x)^2}\left(\psi^n_{k-1}+\psi^n_{k+1}\right).\label{eq:tridiag}
\end{eqnarray}
This is a linear system of equations for the values of $\psi^{n+1}$ on the left-hand side, if the values of $\psi^n$ on the right-hand side are known. Moreover, the form of this system of equations is tridiagonal, which can be solved highly efficiently using the Thomas algorithm \cite{NUMRC}.

In the three-dimensional case, one could proceed with solving for $\psi^{n+1}$ directly, but as the system of equations is no longer tridiagonal, the efficiency related to tridiagonal systems is lost. This is why it is preferable to use the {\em alternating-direction implicit method}, calculating the one-dimensional solution in the $x$, $y$ and $z$ directions, using $\frac{1}{3}\Delta t$ for the time step and $\frac{1}{3}V$ for the potential. This approach is possible because the Hamilton operator can be viewed as a sum of three operators, $\hat{H}=\hat{H}_x+\hat{H}_y+\hat{H}_z$ allowing us to solve numerically using fractional time steps as follows:
\begin{eqnarray}
\hat{H}_x&=&-\frac{1}{2}\frac{\partial^2}{\partial x^2}+\frac{1}{3}V,\nonumber\\
\hat{H}_y&=&-\frac{1}{2}\frac{\partial^2}{\partial y^2}+\frac{1}{3}V,\\
\hat{H}_z&=&-\frac{1}{2}\frac{\partial^2}{\partial z^2}+\frac{1}{3}V.\nonumber
\end{eqnarray}

Substituting these into Eq.~(\ref{eq:tridiag}) (that is, replacing $V$ with $V/3$ and $\Delta x$, respectively, with $\Delta x$, $\Delta y$ or $\Delta z$), using a time step of $\Delta t/3$ we obtain the three fractional iteration steps that correspond to a full iteration with time step $\Delta t$.

As to the nonlinear term, it can be dealt with by a simple iteration that converges rapidly. Specifically, we calculate the non-linear term on the left-hand side by substituting $\phi^n$ in place of $\phi^{n+1}$ and solve the system of equations; we then use this solution to recalculate the non-linear term and solve again, until convergence is obtained. Because in our case, the non-linear term is a cubic term, convergence is very rapid.

\section{Solving Poisson's equation}
\label{sec:PE}

The (non-relativistic) gravitational field corresponding to a distribution of matter characterized by density $\rho$ is given by Poisson's equation for gravity:
\begin{equation}
\nabla^2\phi=4\pi G\rho,
\end{equation}
where $G$ is the gravitational constant. For a BEC, the mass density is given by $\rho=|\psi|^2m$, where $m$ is the mass of the BEC particle. When the BEC condensate is described using the dimensionless Gross-Pitaevskii equation, $m=1$.

A moderately efficient numerical method for solving Poisson's equation is the {\em relaxation method}. This method is based on the finite differences approximation of the second derivative in Poisson's equation:
\begin{eqnarray}
\nabla^2\phi&=&\frac{\partial^2\phi}{\partial x^2}+\frac{\partial^2\phi}{\partial y^2}+\frac{\partial^2\phi}{\partial z^2}\nonumber\\
&=&\frac{\phi(x-\Delta x,y,z)-2\phi(x,y,z)+\phi(x+\Delta x,y,z)}{\Delta x^2}\nonumber\\
&+&\frac{\phi(x,y-\Delta y,z)-2\phi(x,y,z)+\phi(x,y+\Delta y,z)}{\Delta y^2}\nonumber\\
&+&\frac{\phi(x,y,z-\Delta z)-2\phi(x,y,z)+\phi(x,y,z+\Delta z)}{\Delta z^2}+{\cal O}(\Delta^2),
\end{eqnarray}
where $\Delta^2$ is the largest of $\Delta x^2$, $\Delta y^2$ and $\Delta z^2$. If the left-hand side of this equation is given (e.g., by Poisson's equation), this equation can be solved for $\phi(x,y,z)$. The essence of the relaxation method is the realization that these solutions provide successively refined approximations of $\phi(x,y,z)$. In other words, we obtain the iteration formula
\begin{eqnarray}
\phi_{k+1}(x,y,z)&=&\frac{\Delta x^2\Delta y^2\Delta z^2}{2(\Delta x^2\Delta y^2+\Delta y^2\Delta z^2+\Delta z^2\Delta x^2)}\nonumber\\
&&\bigg[\frac{\phi_k(x-\Delta x,y,z)+\phi_k(x+\Delta x,y,z)}{\Delta x^2}\nonumber\\
&+&\frac{\phi_k(x,y-\Delta y,z)+\phi_k(x,y+\Delta y,z)}{\Delta y^2}\nonumber\\
&+&\frac{\phi_k(x,y,z-\Delta z)+\phi_k(x,y,z+\Delta z)}{\Delta z^2}-4\pi G|\psi|^2\bigg].
\end{eqnarray}
This method is accurate but its convergence is slow. However, after the initial configuration of $\phi$ is determined and we iterate the system to the next timestep using the Gross-Pitaevskii equation, $\psi$ and, consequently, $\rho$ will change very little. Therefore, using the values of $\phi^n$ at timestep $n$ as the initial estimate for $\phi^{n+1}$ at timestep $(n+1)$, very rapid convergence is often obtained after just a few iterations.

\section{Choice of units}
\label{sec:UNITS}

In order to put the code presented in this paper to use in an astrophysical context, it is necessary to restore dimensional units.

Use of the dimensionless form of the Gross-Pitaevskii equation amounts to choosing units such that $\hbar=1$ and also $m=1$, where $m$ is the mass of the BEC particle. Given units of length [L], mass [M] and time [T], the choice of $\hbar=1$ amounts to
\begin{equation}
1\frac{([{\rm L}]~{\rm m})^2\cdot([{\rm M}]~{\rm kg})}{([{\rm T}]~{\rm s})}=10^{-34}\frac{{\rm m}^2\cdot{\rm kg}}{\rm s}.
\end{equation}
Conversely, choosing a specific numerical value for $G$ amounts to making the choice
\begin{equation}
G\frac{([{\rm L}]~{\rm m})^3}{([{\rm M}]~{\rm kg})\cdot([{\rm T}]~{\rm s})^2}=6.67\times 10^{-11}\frac{{\rm m}^3}{{\rm kg}\cdot{\rm s}^2}.
\end{equation}
After choosing a specific value for [L], these two equations allow us to determine [M] and [T]:
\begin{eqnarray}
\left[{\rm T}\right]&=&5.3\times 10^{14}[{\rm L}]^{5/3}\sqrt[3]{G}~{\rm s},\\
\left[{\rm M}\right]&=&5.3\times 10^{-20}[{\rm L}]^{-1/3}\sqrt[3]{G}~{\rm kg}.
\end{eqnarray}
Specifically, if we choose our unit of length to be $1$~kpc~$\simeq 3\times 10^{19}$~m, we get
\begin{eqnarray}
\left[{\rm T}\right]&=&1.5\times 10^{47}\sqrt[3]{G}~{\rm s},\\
\left[{\rm M}\right]&=&1.7\times 10^{-26}\sqrt[3]{G}~{\rm kg}\simeq 10\sqrt[3]{G}~{\rm GeV}.
\end{eqnarray}

Choosing $G=10^{-100}$ yields the units
\begin{eqnarray}
\left[{\rm T}\right]&=&7\times 10^{13}~{\rm s}\simeq 2.2\times 10^6~{\rm year},\\
\left[{\rm M}\right]&=&8\times 10^{-60}~{\rm kg},
\end{eqnarray}
corresponding to a BEC particle mass of $4.4\times 10^{-24}$~eV.

Finally, velocity is measured in units of kpc/(2.2 million years)$~\simeq 430$~km/s.

\section{Initial and boundary conditions}
\label{sec:INIT}

Numerically solving a system of coupled differential equations requires a set of initial and boundary conditions.

Specifically, solving the Gross-Pitaevskii equation requires an initial field configuration $\psi(x,y,z)$ at $t=0$. In turn, the numerical solution of Poisson's equation for gravity needs boundary conditions in the form of values $\phi(x_{\rm min},y,z)$, $\phi(x_{\rm max},y,z)$, $\phi(x,y_{\rm min},z)$, $\phi(x,y_{\rm max},z)$, $\phi(x,y,z_{\rm min})$ and $\phi(x,y,z_{\rm max})$.

For the purpose of testing our code, we chose a very simple initial density profile, related to solutions of the Lane-Emden equation, centered around the origin at $x=0,y=0,z=0$:
\begin{equation}
\rho=\rho_0(R^2-r^2),
\end{equation}
where $r^2=x^2+y^2+z^2$ and $R$ is a characteristic radius. Given $\rho$, we calculate $\psi=\sqrt{\rho}$, making the initial value of $\psi$ purely real everywhere.

We emphasize that this choice does not necessarily represent a physically viable configuration; it was strictly used for code testing and validation. Application of the code involves, among other things, choosing initial density profiles that reflect valid physical assumptions. For instance, one may opt to use a Navarro-Frank-White distribution \cite{NFW1997} to model the initial halo density at $t=0$.

As to the boundary condition, we simply assume that the gravitational field vanishes on the boundaries of the simulation volume. Since in the context of Newtonian gravity, the gravitational field is indeterminate up to an additive constant, this amounts to the assumption the gravitational field is constant on the boundary. While this is somewhat artificial, one may justify this choice by noting that real galaxies exist in an external gravitational field that in turn is determined by other, more distant galaxies and clusters; this external field is imposed upon the dynamics of a real, physical galaxy the same way we impose our boundary condition on the model galaxy. In any case, if the simulation volume is sufficiently large compared to the galaxy being simulated, the geometry of the boundary will play no significant role in the model galaxy's evolution.

\section{Implementation and testing}
\label{sec:CODE}

Our code was originally implemented in FORTRAN 90. Later, however, we decided to port the code to C++, which resulted in a twofold performance improvement.

We also experimented with a GPGPU\footnote{General Purpose Graphics Processing Unit} accelerated version that was designed to run using the OPENCL library, on an AMD 6900-class graphics card. This version yielded a considerable improvement in performance, but only when single-precision arithmetic was used. The magnitude of the quantities needed in the astrophysical context (e.g., $G=10^{-100}$) precluded the use of single-precision floats. The performance improvement of the GPGPU version was not sufficient to justify further rewriting the code; therefore, the GPGPU implementation was, for the time being, abandoned, although the code remains functional.

The C++ version of the code was heavily tested using various grid sizes. When testing the code, we took to heart the important advice offered by the authors of Ref. \cite{NUMRC}: ``{\em You should always first run your programs on {\rm very small} grids, e.g., $8\times 8$, even though the resulting accuracy is [...] poor [...] [N]ew instabilities sometimes do show up on {\rm larger} grids, but old instabilities never (in our experience) just go away.}'' We determined that our code runs very well on a grid size of $20\times 20\times 20$, and the simulation is very rapid; this makes it easy to select physically interesting test cases that can be further analyzed using a much finer grid. On the other hand, we found that even with a grid size of $120\times 120\times 120$, a simulation that models the evolution over $10^{9}$~years can be completed in the course of a day or so on modern desktop computer hardware.

We specifically tested the numerical robustness of our code by verifying that the wavefunction $\psi$ remains unitary. Even after 50,000 iterations, $\int_V|\psi|^2~dV$, when evaluated for the entire simulation volume, remained within a few percent of its initial value.

\begin{figure}[c]
\caption{Illustrative example of the evolution of a rotating, self-gravitating BEC, in a $100\times 100\times 100$~kpc$^3$ volume, simulated using an $80\times 80\times 80$ spatial grid. The calculated rotational velocity and density are shown after the 1, 250 and 500 iteration time units, the latter corresponding to $\sim 1$~Gyr. The total mass of this condensate is $\sim 6.8\times 10^{12}~M_\odot$.\label{fig:sim500}}
\vskip 1em
\newcolumntype{V}{>{\centering\arraybackslash} m{.15\linewidth} }
\begin{tabular}{V|ccc}
Iteration:&1&250&500\\\hline
~&~&~&~\\
\vskip -8em $v_{\rm rot}$&\includegraphics[width=0.25\linewidth]{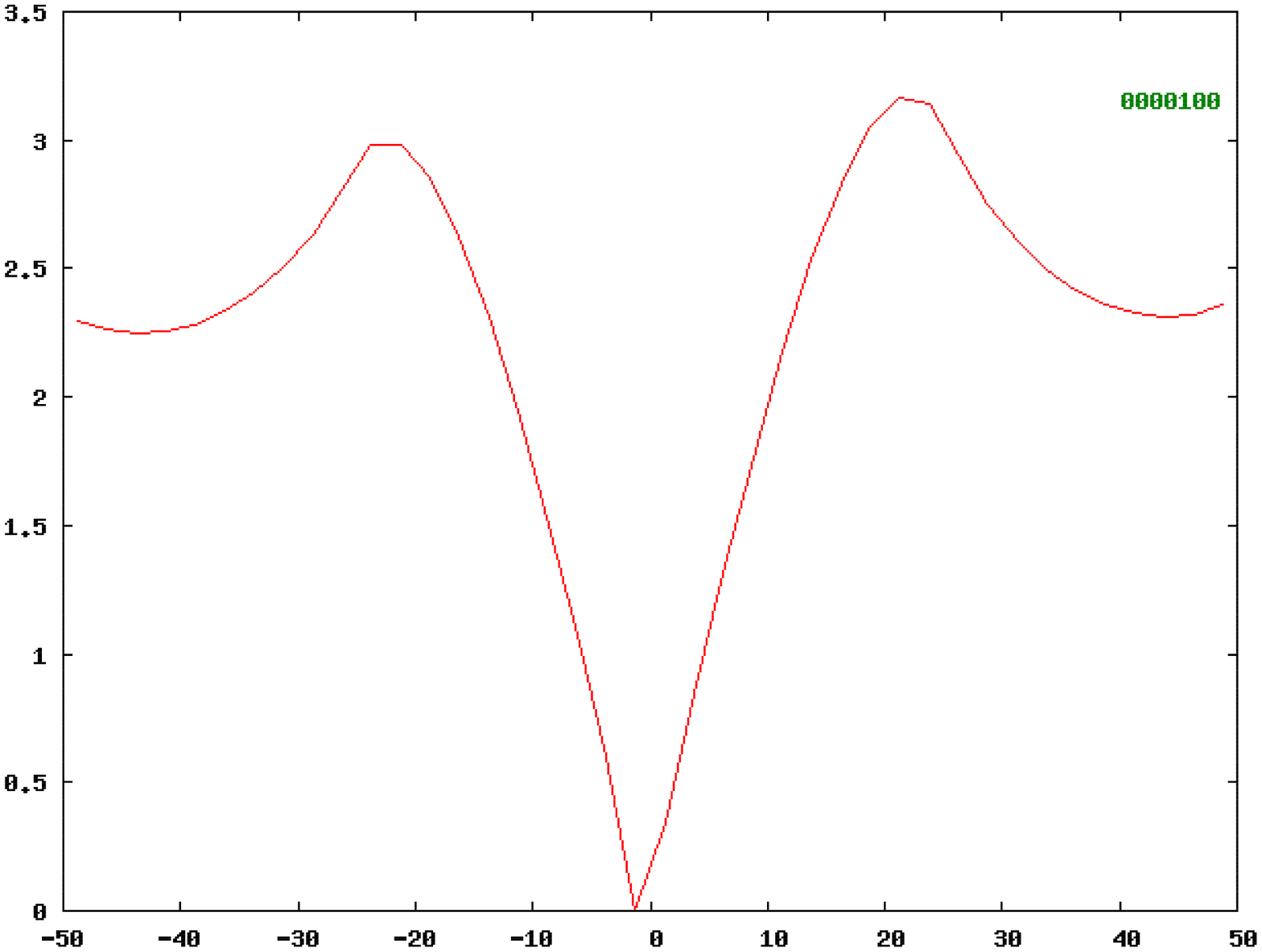}&\includegraphics[width=0.25\linewidth]{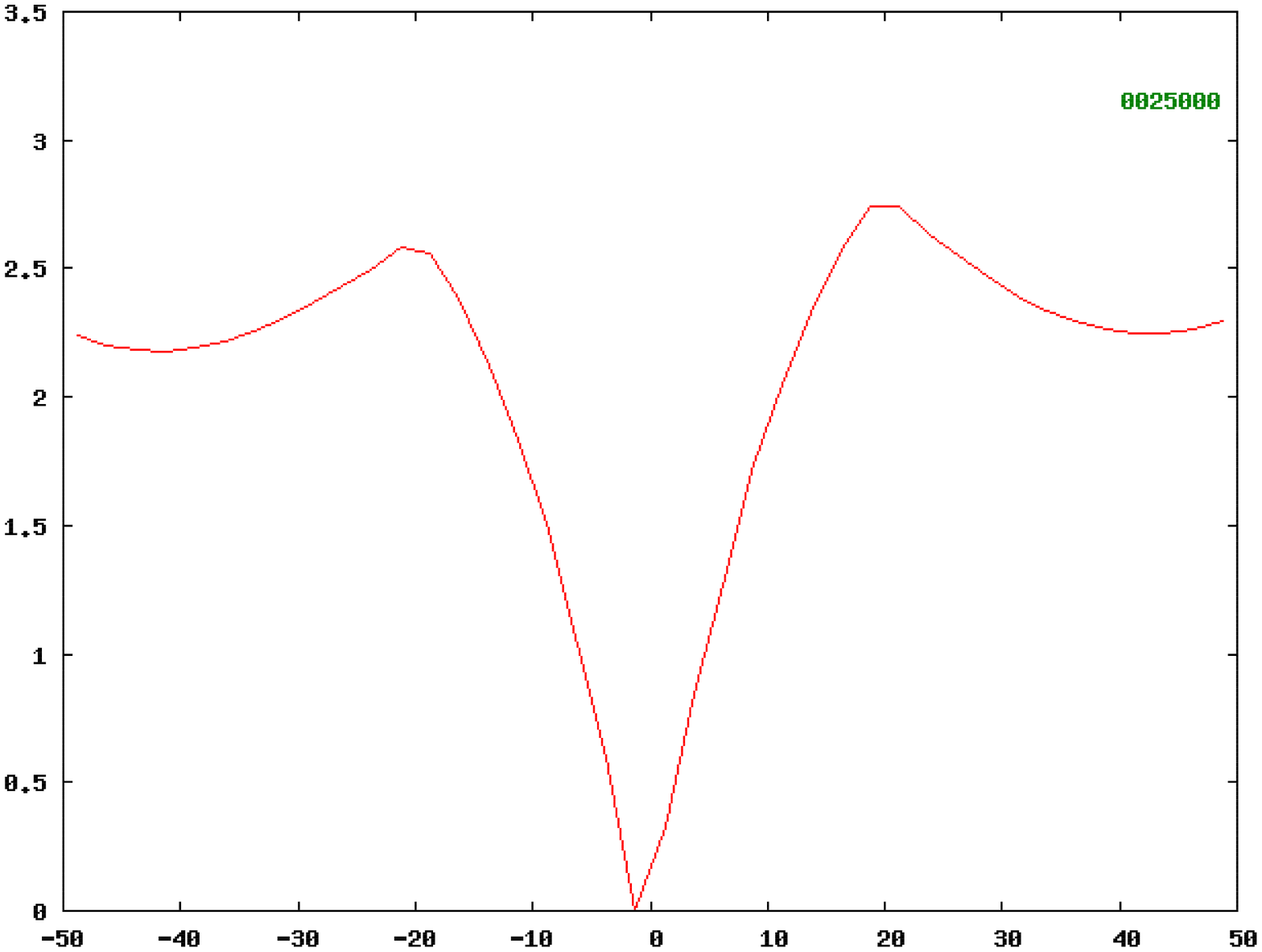}&\includegraphics[width=0.25\linewidth]{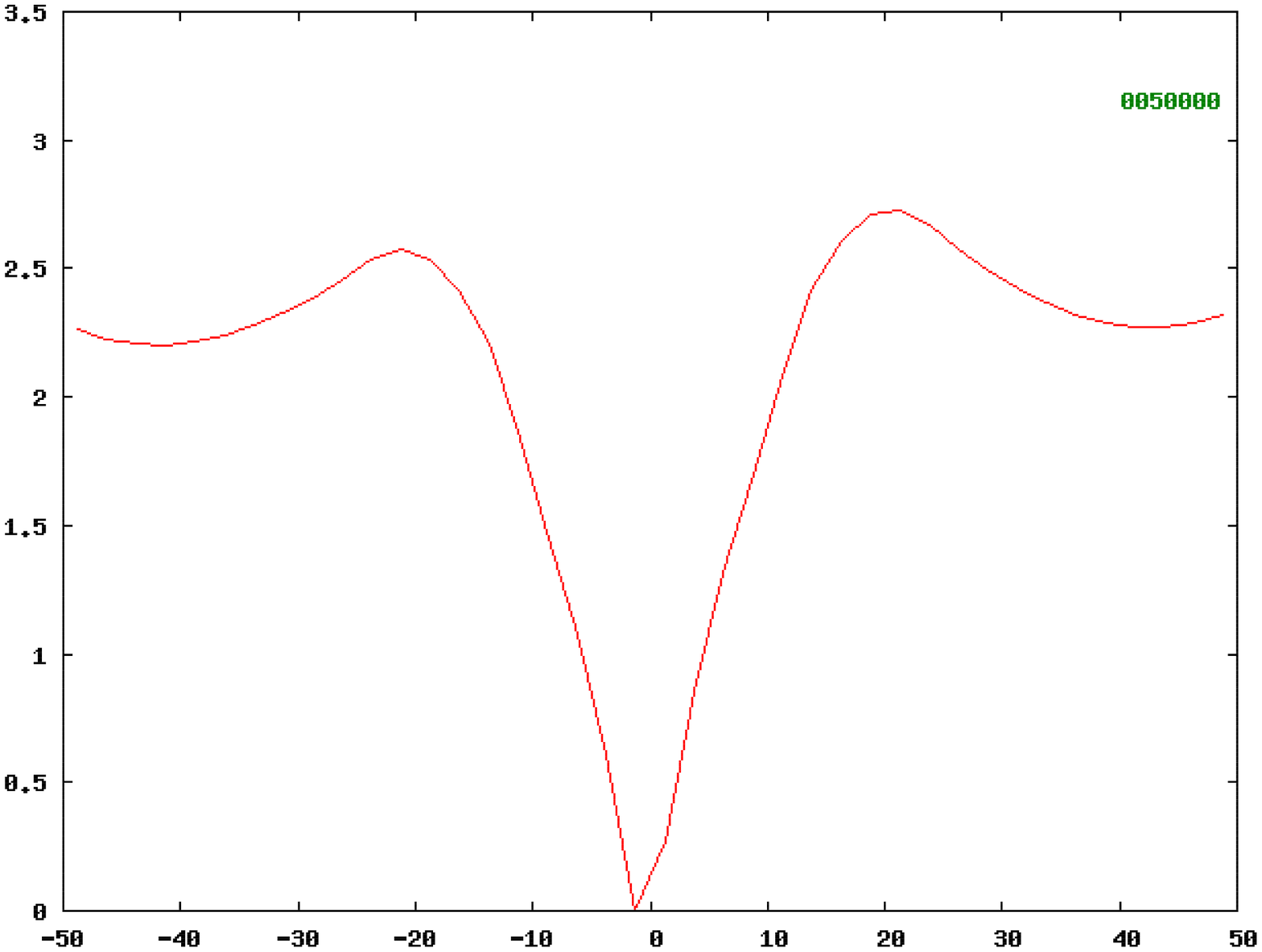}\\\hline
~&~&~&~\\
\vskip -8em $|\psi|^2$ in $xy$-plane&\includegraphics[width=0.25\linewidth]{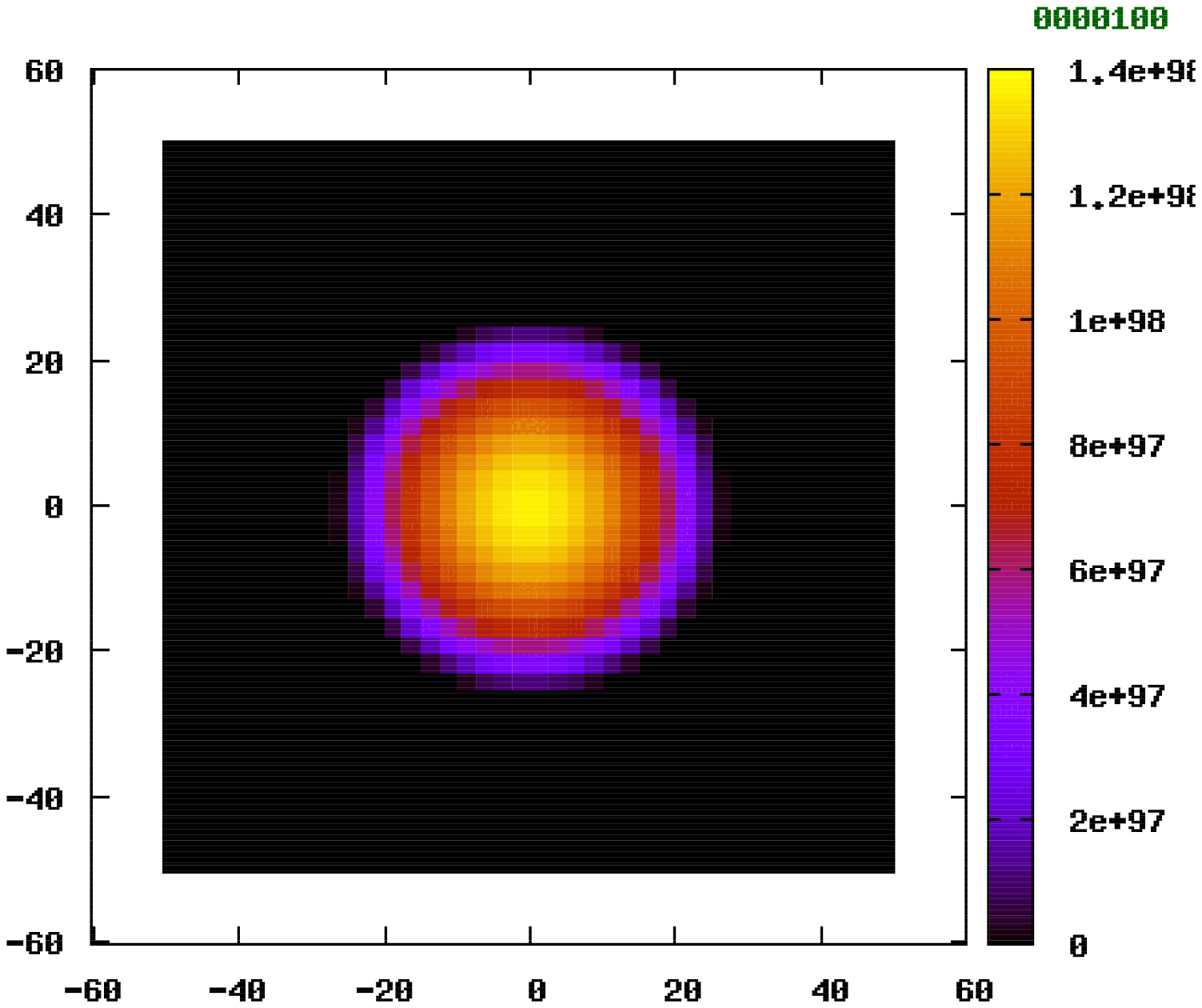}&\includegraphics[width=0.25\linewidth]{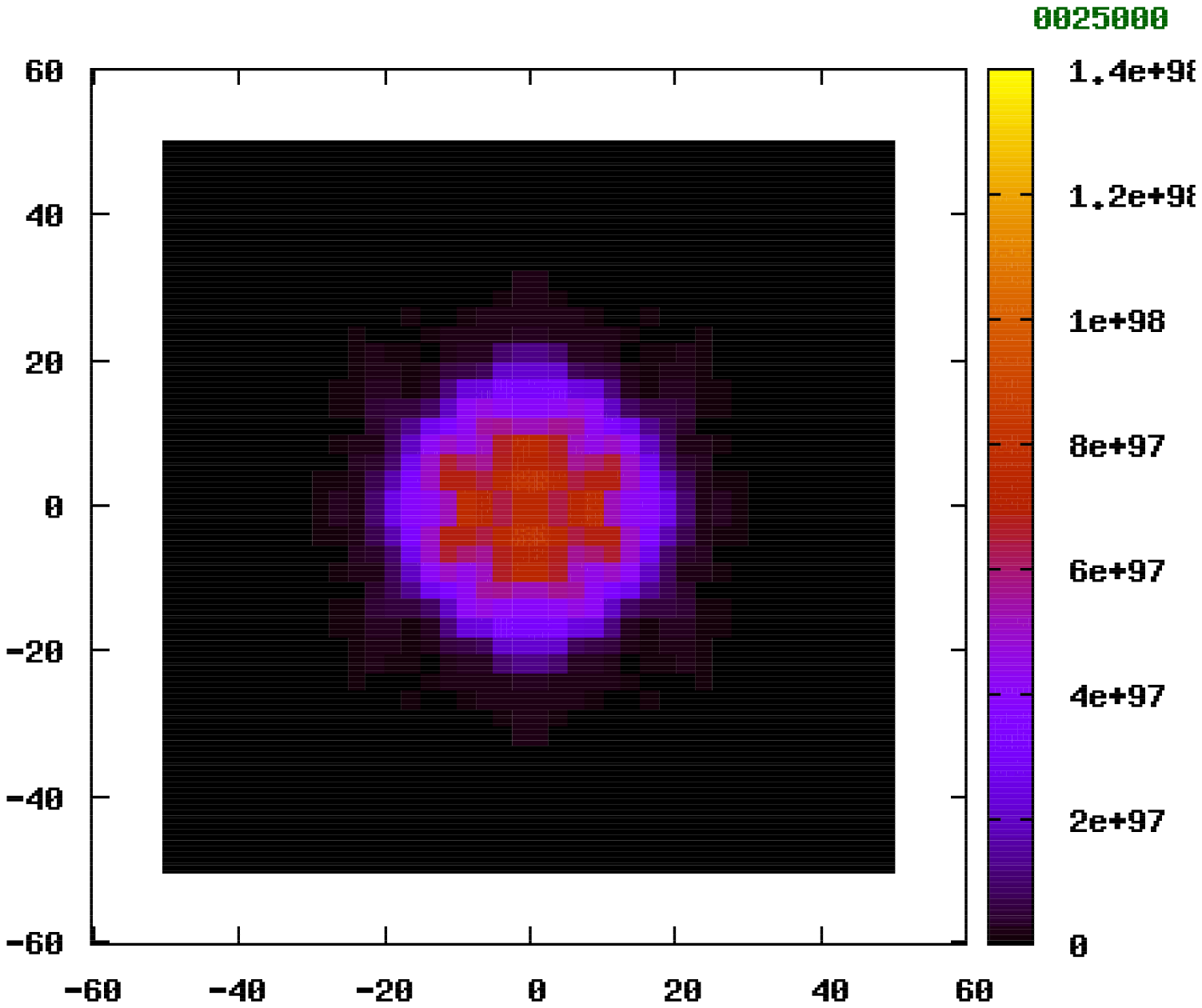}&\includegraphics[width=0.25\linewidth]{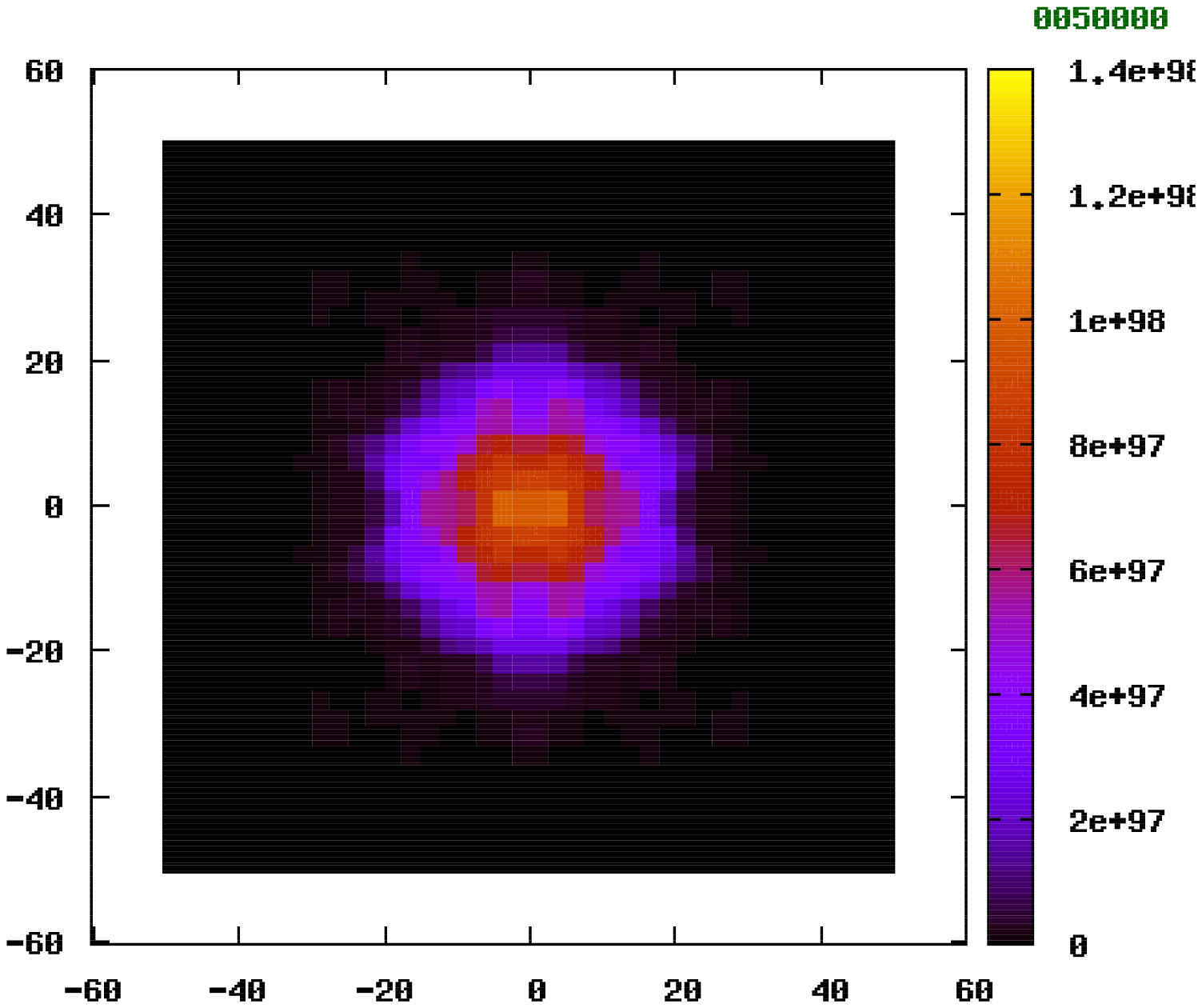}\\\hline
~&~&~&~\\
\vskip -8em $|\psi|^2$ in $yz$-plane&\includegraphics[width=0.25\linewidth]{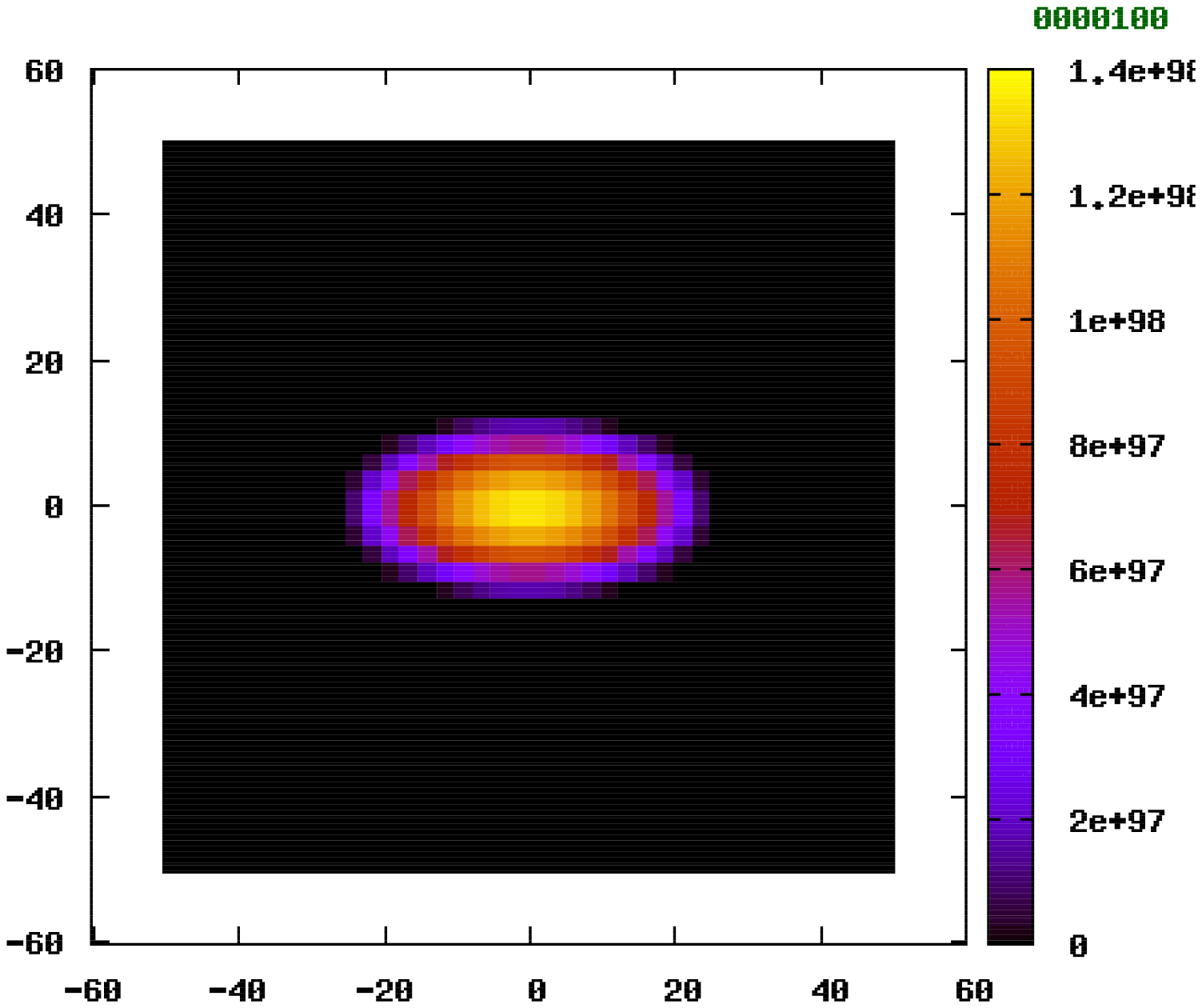}&\includegraphics[width=0.25\linewidth]{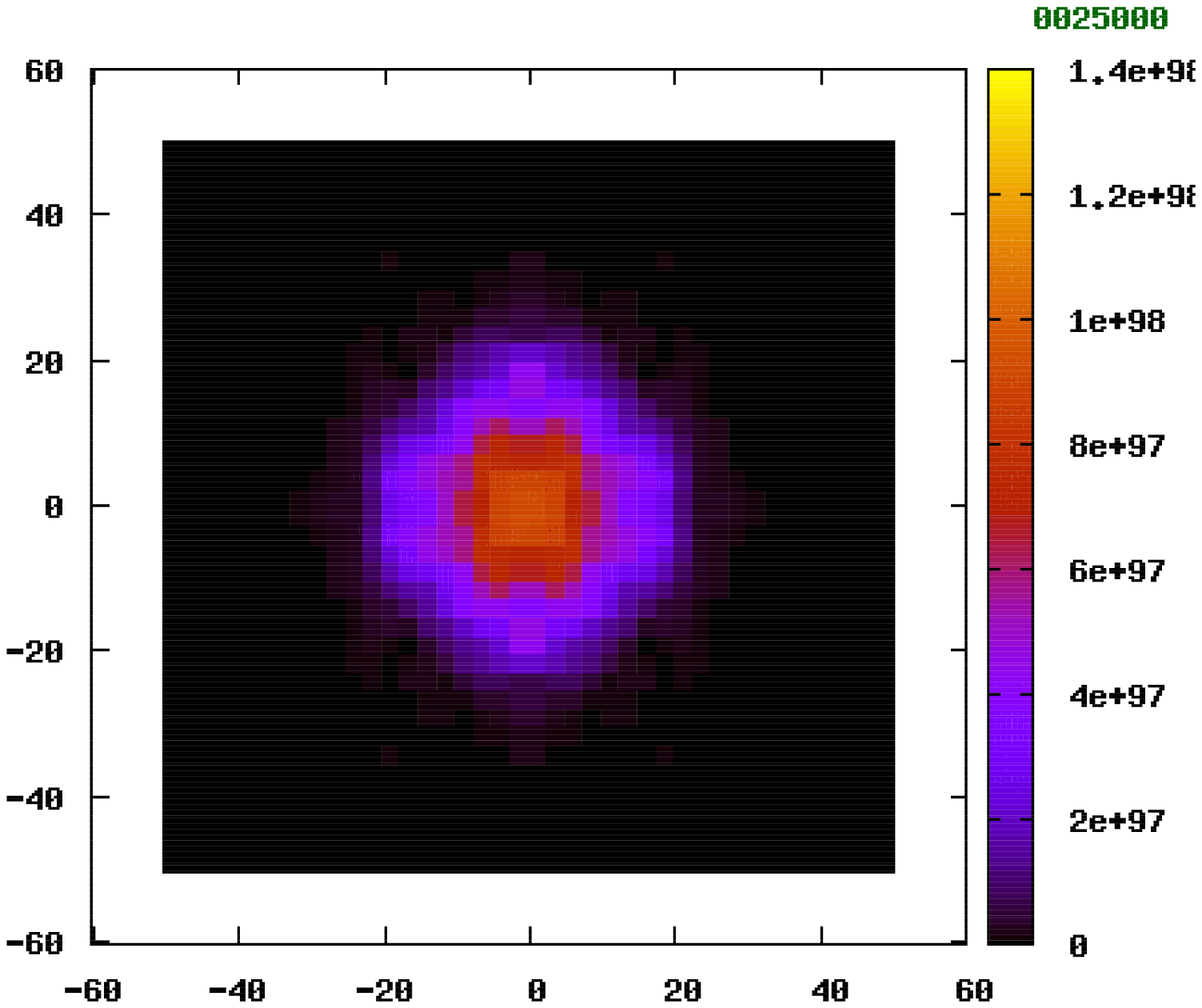}&\includegraphics[width=0.25\linewidth]{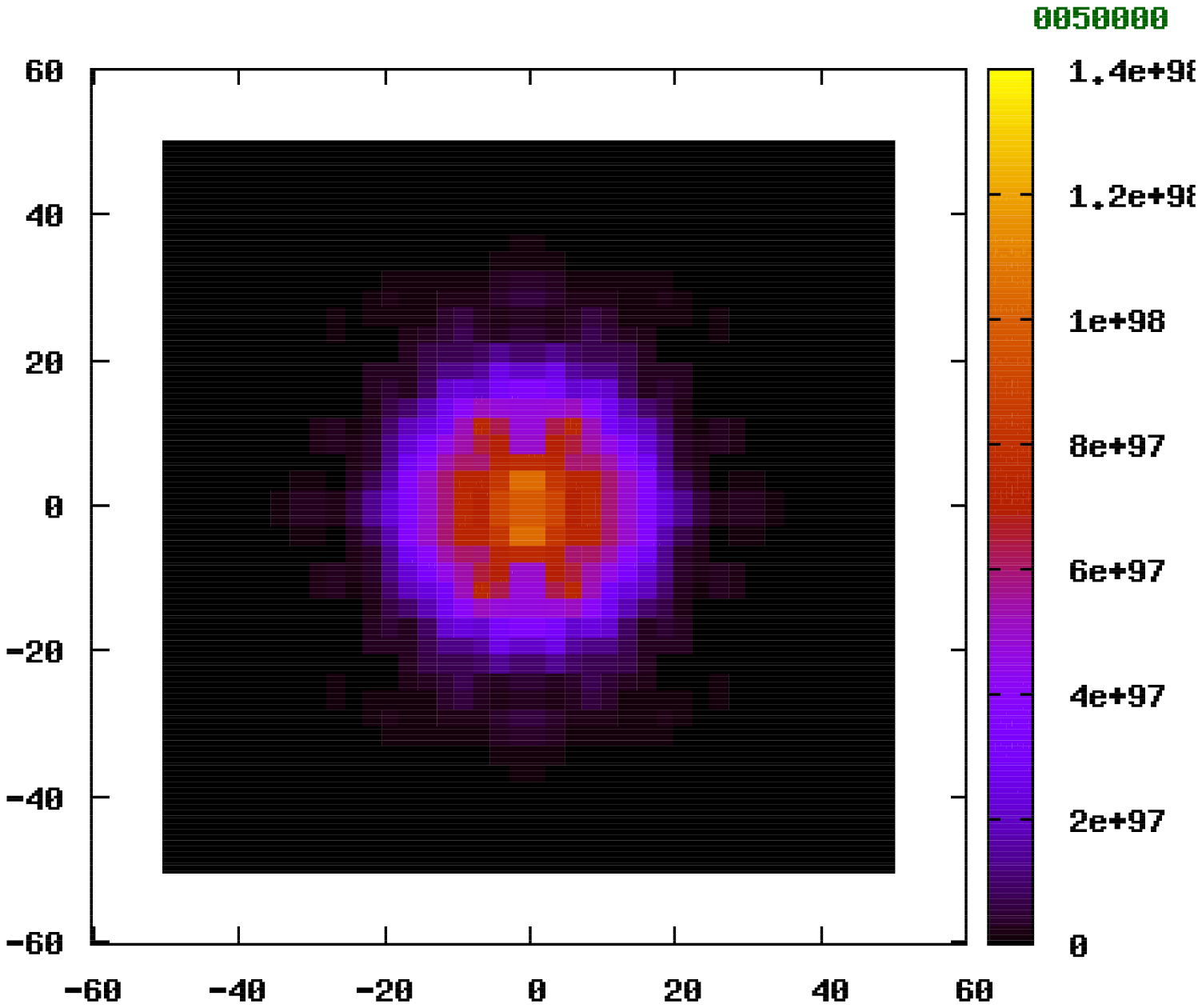}
\end{tabular}
\end{figure}

Specifically, we ran a test simulation in a $100\times 100\times 100$~kpc$^3$ volume, using an $80\times 80\times 80$ cells, using a time step of $\Delta t=0.01$ units of time (corresponding to $\sim 22,000$ years.) For the entire volume, $\int~|\psi|^2~dV=1.7\times 10^{102}$, corresponding to a total BEC mass of  $\sim 6.8\times 10^{12}~M_\odot$. The total simulation time of 50,000 iterations corresponds to $\sim 1.1\times 10^9$~years; the results of this simulation are shown in Fig.~\ref{fig:sim500}.

\section{Conclusions}
\label{sec:END}

Development of the software code described in this paper is now complete, with the code yielding expected results for test cases. The next step is to find suitable initial conditions to model a BEC dark matter halo that may surround a real galaxy. The question is whether halo configurations can be found that are stable over time scales of $10^{10}$ years, and yield circular orbital velocities that remain approximately constant at different radii.

Results of this on-going investigation will be reported when they become available.

\section*{Acknowledgements}

EJMM thanks Profs. C.~F. Barenghi and M. Tsubota for help with the development of the first version of the code presented in this manuscript.





\bibliographystyle{model1-num-names}
\bibliography{refs}







\end{document}